\documentclass[journal=apchd5,manuscript=article,layout=traditional]{achemso}
\pdfoutput=1
\usepackage[version=3]{mhchem} % Formula subscripts using \ce{}
\usepackage{amssymb}
\usepackage{verbatim}
\usepackage{color,xcolor}
\newcommand{\ii} {\mathrm{i}}
\newcommand{\ee} {\mathrm{e}}
\newcommand{\vep}{\varepsilon}
\newcommand{\vect}[1] {\boldsymbol{\mathbf{#1}}}
\newcommand{\vers}[1] {\boldsymbol{\mathbf{\hat{#1}}}}
\newcommand{\abs}[1] {\mathopen{}\left|#1\right|\mathclose{}}
\newcommand{\sqabs}[1] {\mathopen{}\left|#1\right|^2\mathclose{}}
\newcommand{\ccpar}[1] {\mathopen{}\left(#1\right)\mathclose{}}
\newcommand{\sqpar}[1] {\mathopen{}\left[#1\right]\mathclose{}}
\newcommand{\clpar}[1] {\mathopen{}\left\{#1\right\}\mathclose{}}

\newcommand{\dd}[1] { \mathrm{d}#1\mbox{ }}
\newcommand{\mat}[1] {\begin{bmatrix} #1 \end{bmatrix}}
\newcommand{\unit}[1] {\mbox{ }\mathrm{#1}}

\DeclareMathOperator{\arccosh}{arccosh}
\DeclareMathOperator{\sign}{sgn}

\author{Eduardo J. C. Dias}
\email{eduardo.dias@fisica.uminho.pt}
\author{N. M. R. Peres}
\email{peres@fisica.uminho.pt}
\affiliation[Department of Physics and Center of Physics, University of Minho]{Department of Physics and Center of Physics, and QuantaLab, University of Minho, 4710--057, Braga, Portugal}

\title[An \textsf{achemso} demo]
  {Controlling spoof plasmons in a metal grating using graphene surface plasmons}

\abbreviations{EM}

\begin{document}
%%%%%%%%%%%%%%%%%%%%%%%%%%%%%%%%%%%%%%%%%%%%%%%%%%%%%%%%%%%%%%%%%%%%%
%% The manuscript does not need to include \maketitle, which is
%% executed automatically.  The document should begin with an
%% abstract, if appropriate.  If one is given and should not be, the
%% contents will be gobbled.
%%%%%%%%%%%%%%%%%%%%%%%%%%%%%%%%%%%%%%%%%%%%%%%%%%%%%%%%%%%%%%%%%%%%%
\begin{abstract}
  Spoof plasmons mimic noble metal plasmons. The equivalent of the plasma frequency is an energy scale imposed 
  by the geometry of the metal grating upon which they propagate. In this paper we show that the dispersion of
  spoof plasmons in the THz can be controlled placing a doped graphene sheet on the proximity of a metallic grating, adding more versatility to this type of system. 
  We develop a semi-analytical model, based on a perfect-metal diffraction grating. This model allows to reproduce
  well FDTD calculations for the same problem but with much less computer time. We discuss the optical properties of the 
  system covering a spectral range spanning the interval from the THz to the mid-IR. It is shown that the system can be used
  as both a perfect absorber and a sensing device. For illustrating the latter property we have chosen different 
  alcohols as analytes. The frequency at which perfect absorption appears can be controlled by the 
  geometric parameters of the grating and by the value of the Fermi energy in graphene. The theoretical results predicted
  throughout this work can be verified experimentally in the future.
  
  %Within these applications, this
  %paper makes several theoretical predictions that can be verified experimentally.
\end{abstract}

\section{Introduction}

It is a well known result that a perfect flat conducting surface does not support surface plasmon polaritons (SPPs)\cite{maradudin2014modern}. A real metal/dielectric interface, on the other hand, does support SPPs, but their energy is of the order of the plasma energy of the metal, which, for most good plasmonic metals (i.e. with relatively low losses, like gold or silver) is in the ultraviolet spectral range\cite{ordal1985optical}, which is an energy scale too high for many sensing applications.

An usual method to excite low-energy plasmons in a perfect metal is to introduce periodic structures (namely grooves or holes) in its surface. Under these conditions, the corrugated system supports surface modes which, strictly speaking, are not SPPs (because those are excited in flat interfaces), but effectively mimic their properties, and for that reason they are usually referred to as \textit{spoof plasmons} (or spoof surface plasmons, SSPs). These plasmons were firstly studied and named by \citeauthor{pendry2004mimicking} in \citeyear{pendry2004mimicking}, who showed that these solutions were equivalent to the ones retrieved from a flat metal/dielectric interface with the permittivity of the metal being modelled by an effective permittivity dependent on the geometry of the grooves\cite{pendry2004mimicking}. Using that method, the authors were able to calculate an approximate dispersion relation for these plasmons. Later, this method was generalized to show that spoof plasmons are supported by real periodically-grooved metals\cite{rusina2010theory}, and different authors have also suggested that spoof plasmons can be excited in small structures with only a few unit cells\cite{joy2017spoof}.

The advantage of this type of plasmons is that their energy depends highly on the geometry of the grooves, and particularly on their depth. In fact, using the effective permittivity method, one finds that the effective plasma energy of the corrugated perfect metal ---which allows the estimation of the spoof plasmons energy--- is given by\cite{erementchouk2016electrodynamics}
\begin{equation}
 \omega^{\ce{eff}}_{\ce{p}} = \dfrac{\pi c}{2 h \sqrt{\vep}},
 \label{eq:plasmafreq}
\end{equation} 
where $h$ is the depth of the grooves and $\vep$ the permittivity of the medium inside the grooves. This result (which will be rederived in this work, using a different method) suggests that, in general, the scale of the grooves' dimensions has a substantial impact on the frequency of the spoof plasmons. For this reason, these plasmons have been regarded recently as a promising alternative to traditional SPPs for applications that include waveguides\cite{vogt2016plasmonic,tian2016compact,huang2016metal,mousavi2010highly}, couplers/decouplers\cite{xu2016decoupling,zhang2016three}, leaky wave antennas\cite{kianinejad2017single,panaretos2016sinusoidally,lu2016periodic} and sensing devices\cite{ng2013spoof,ng2014broadband,yao2014high}, specially in the THz spectral range (which is very useful, since many fundamental excitations ---like phonons in a lattice or molecule vibrations in a gas--- have frequencies which lie in the THz and mid-IR spectral regions\cite{mittleman2013frontiers}). %It is in this application that the present work will mainly focus.

%The usage of spoof plasmons for sensing purposes has already been proposed previously by \citeauthor{ng2013spoof}\cite{ng2013spoof}, who were able to measure the dispersion relation of these plasmons, using the scattering edge coupling method, for several different fluid placed inside the grooves.

Although the optimization of the geometry of the grooves allows an effective tuning of the frequency of the spoof plasmons, it has the disadvantage of being immutable for some certain system ---and many times, building many different systems with slightly different geometric parameters may not be the most practical solution. In traditional plasmonics, the usual solution for this limitation is the introduction of graphene, because its Fermi energy (that strongly controls its plasmonic response\cite{goncalves2016introduction}) can be easily varied either through chemical doping or the application of a gate potential\cite{wang2008gate}. Other advantages of graphene include its high carrier mobility (what translates in small losses, compared to noble metals\cite{jablan2009plasmonics}) and a strong light confinement in its surface\cite{koppens2011graphene}. 

Following these features, we propose in this work the usage of a doped graphene sheet to effectively tune the plasmonic properties of the spoof plasmons, by controlling graphene's Fermi energy and the distance to the grooved metallic surface. A first approach to this problem has been carried out by \citeauthor{ding2015coupling}\cite{ding2015coupling}, for a 2D metal grating, using the effective medium approach to describe the periodic region of the structure; we, on the other hand, consider a 1D metal grating, and have employed a mode-matching method considering the metal to be perfect, what will prove to be a good approximation. Moreover, we have accounted for the non-local effects in the graphene conductivity, using Mermin's formula (see the Supporting Information for further details). The addition of graphene is shown to be very efficient when the energy of the uncoated diffraction grating SSPs is close to the energy of the graphene plasmons, especially in the THz spectral range. Afterwards, we will propose how this tunnability can be used for waveguiding and sensing applications.

%This can be readily shown when considering a semi-infinite metallic medium comprising the region $y<0$ and a dielectric with permittivity $\vep$ comprising the region $y>0$. In the latter region, Maxwell's equations allow, in general, p-polarized solutions with the form $\vect{B}\ccpar{\vect{r},t} = \ee^{\ii (q_x x + q_z z)} \ccpar{ \alpha^+ \ee^{\kappa y} + \alpha^- \ee^{-\kappa y} } \ee^{-\ii \omega t} \vers{z}$, where $\omega$ is the frequency of the plasmons, $q=(q_x^2+q_z^2)^{1/2}$ is their in-plane momentum, $\kappa$ is their out-of-plane momentum (fixed by $q$ and $\omega$) and $\alpha^{\pm}$ are undetermined coefficients fixed by the boundary conditions of this region. Since the magnetic field must not diverge far away from the metal ($y \to \infty$), the coefficient $\alpha^+$ must be equal to $0$.  On the other hand, in the region $y<0$ (perfect metal), no electric fields are allowed, and therefore the tangential electric field (which is continuous across any interface) must vanish at $y=0$; using this condition, one easily finds that $\alpha^-$ must also vanish. Since the only solution that obeys both boundary conditions implies that $\alpha^+=\alpha^-=0$, we conclude that these plasmonic modes are not supported by a flat perfect conducting interface.

\section{Spoof plasmons in a grooved metallic surface coated with graphene}
\label{sec:math}

\begin{figure}[htb]
  \includegraphics[width=0.75\textwidth]{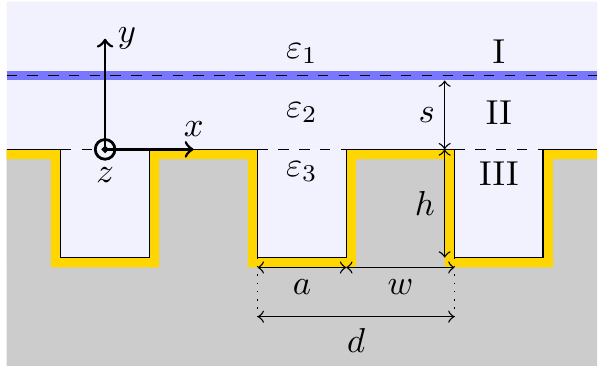}
  \caption{Front-view schematic representation of the system presently under study. The grey area depicts a substract where the the perfect metal thin film, in gold color, is deposited. The light-blue areas depict the dielectric regions and the dark-blue thin area is the graphene sheet. All the different geometrical parameters and dielectric functions in either region I--III are marked in the figure. The system is assumed to be infinite in the $z$-direction and periodic in the $x$-direction.}
  \label{fig:Scheme}
\end{figure}

%In this work, we propose a way to take advantage of graphene's remarkable tunability in order to further improve spoof plasmon-based sensors. For that purpose, 
Let us consider a semi-infinite grooved perfect-conducting surface coated by a graphene sheet parallel to its surface, like the one depicted in Figure \ref{fig:Scheme}.

We divide the dielectric area in three different regions (I, II and III) which can, in general, have different permittivities ($\vep_1$, $\vep_2$ and $\vep_3$, respectively). The grooves in the metal are assumed to be rectangular, with width $a$, depth $h$, and period $d$; the metal itself is assumed to be perfect. The graphene sheet is lying at a distance $s$ from the top of the grooves, and is described by a non-local conductivity $\sigma$ (which will be discussed later) dependant on its Fermi energy $E_{{\ce{F}}}$ and relaxation energy $\Gamma$. 

\subsection{Dispersion Relation}
\label{sec:math:DR}

The procedure we adopted to characterize the spoof plasmons consists on the modal decomposition of their fields in a Fourier series. This method has been used in the literature before\cite{maradudin2014modern,maradudin2016rayleigh,shen2008terahertz}, but never in the presence of graphene. The starting point is the solution of the wave equation retrieved from Maxwell's equations in either region I, II or III. Considering explicitly p-polarized modes with a frequency $\omega$ and a harmonic time-variation $\ee^{-\ii \omega t}$, the magnetic field in region $\lambda$ must have the form $\vect{B}_   {\ce{\lambda}}\ccpar{\vect{r},t} = B_   {\ce{\lambda}}\ccpar{x,y} \ee^{-\ii \omega t} \vers{z}$ (omitting the $z$-dependence) with
\begin{equation}
 B_   {\ce{l}} \ccpar{x,y} = \sum_{n=-\infty}^{\infty} B_n \ee^{\ii \beta_n x} \ee^{\ii \kappa^{(1)}_n y},
 \label{eq:BI}
\end{equation} 
\begin{equation}
 B_   {\ce{ll}} \ccpar{x,y} = \sum_{n=-\infty}^{\infty} \ee^{\ii \beta_n x} \sqpar{ C_n^+ \ee^{\ii \kappa^{(2)}_n y} + C_n^- \ee^{-\ii \kappa^{(2)}_n y} },
 \label{eq:BII}
\end{equation} 
\begin{equation}
 B_   {\ce{lll}} \ccpar{x,y} = \sum_{n=0}^{\infty} A_n \cos\sqpar{ \frac{n \pi}{a} \ccpar{ x - \frac{a}{2} } } \cos\sqpar{ \kappa^{(3)}_n \ccpar{ y + h } },
 \label{eq:BIII}
\end{equation} 
and the respective electric fields are given by 
\begin{equation}
 \vect{E}_   {\ce{\lambda}}(\vect{r},t) = \ccpar{\frac{\ii c^2}{\omega \vep_   {\ce{\lambda}}}} \vect{\nabla} \times \vect{B}_   {\ce{\lambda}}(\vect{r},t).
 \label{eq:EfromB}
\end{equation}
Note that equations \eqref{eq:BI}--\eqref{eq:BIII} were written explicitly to ensure that (i) the fields in region I do not diverge as $y \to \infty$; and (ii) the tangential component of the electric field in region III always vanishes in the surface of the metal. This is a consequence of the fact that a perfect metal does not admit non-null electric fields on its inside, and the tangential electric field is continuous across any interface. The coefficients $A_n$, $B_n$ and $C^{\pm}_n$ are, for the moment, unknown, while the electromagnetic wave equation imposes that $\kappa^{(1)}_n = \kappa\ccpar{\beta_n,\vep_1}$, $\kappa^{(2)}_n = \kappa\ccpar{\beta_n,\vep_2}$ and $\kappa^{(3)}_n = \kappa\ccpar{n \pi/a,\vep_3}$, with $\kappa\ccpar{q,\vep} \equiv \sqrt{\vep \omega^2/c^2-q^2}$. Furthermore, since the grooved system is periodic, the corresponding fields must obey the Bloch's Theorem, $\vect{B}\ccpar{\vect{r}+d \vers{x}} = \ee^{\ii q d} \vect{B}\ccpar{\vect{r}}$, which means that $\beta_n = q + 2n\pi/d$ ($q$ being the momentum of the plasmons in the $x$-direction). Another consequence of Bloch's Theorem is that we need only to determine the fields in an unit cell $\abs{x}<d/2$ of the system, and the fields elsewhere are then totally determined.

In order to find the coefficients $A_n$, $B_n$ and $C^{\pm}_n$, we need to evaluate the boundary conditions at the interfaces I/II and II/III. The detailed derivation from this point onwards is presented in the Supporting Information (SI), where we show that the dispersion relation of the spoof plasmons in this system is given by the matrix equation $\det(\mathbb{M}-\mathbb{I}) = 0$, where $\mathbb{I}$ is the unit matrix with elements $\sqpar{\mathbb{I}}_{\ell m} = \delta_{\ell m}$ and $\mathbb{M}$ is a matrix whose elements $\sqpar{\mathbb{M}}_{\ell m} = M_{\ell m}$ are given by
\begin{equation}
 M_{\ell m} = \ii \frac{\vep_2}{\vep_3} \frac{a}{d} \ccpar{ \frac{2}{1+\delta_{\ell 0}} } \sum_{n=-\infty}^{\infty} \ccpar{ \frac{\chi_n^+ + \chi_n^-}{\chi_n^+ - \chi_n^-} } \frac{\kappa^{(3)}_m}{\kappa^{(2)}_n} \frac{\sin\ccpar{\kappa^{(3)}_m h}}{\cos\ccpar{ \kappa^{(3)}_{\ell} h }} S^{\ast}_{n \ell} S_{nm},
\end{equation} 
with
\begin{equation}
  \chi_n^{\pm} = \frac{1}{2} \sqpar{ 1 + \frac{ \sigma\ccpar{\beta_n, \omega} \kappa_n^{(1)}}{\omega \vep_0 \vep_1} \pm \frac{\vep_2 \kappa_n^{(1)}}{\vep_1 \kappa_n^{(2)}} } \ee^{\ii \kappa_n^{(1)} s} \ee^{ \mp \ii \kappa_n^{(2)} s}
\end{equation} 
and $S_{\ell n} = \frac{1}{a} \int_{-a/2}^{a/2} \dd{x} \ee^{-\ii \beta_{\ell} x} \cos\sqpar{\frac{n \pi}{a}\ccpar{x-\frac{a}{2}}}$ in an integral with an analytical solution. In the previous expression, $\sigma\ccpar{q, \omega}$ is the non-local conductivity of the graphene sheet (consult the SI for further details); note therefore that all the influence of the graphene in the dispersion is contained in the factors $(\chi^+_n + \chi^-_n)/(\chi^+_n - \chi^-_n)$ in each term of the sum, which are equal to $1$ in its absence.

At this point, it is interesting to observe that a simple approximation we could have done to simplify our calculations was to consider that, in region III (the grooves), only the lowest mode $n=0$ was non-zero (that is, $A_n=A_0 \delta_{n0}$). In doing so, the corresponding equation for the dispersion relation of the spoof plasmons becomes
\begin{equation}
 1=\ii \frac{\vep_2}{\vep_3} \frac{a}{d} \frac{\sqrt{\vep_3} \omega}{c} \tan\ccpar{\frac{\sqrt{\vep_3} \omega}{c} h} \sum_{n=-\infty}^{\infty} \ccpar{ \frac{\chi_n^+ + \chi_n^-}{\chi_n^+ - \chi_n^-} } \frac{\abs{S_{n 0}}^2}{\kappa^{(2)}_n},
 \label{eq:approxDR}
\end{equation} 
where we have used the explicit expression for $\kappa^{(3)}_0$. The first thing to note in the above expression is that, in the absence of graphene and dispersive dielectric media, it can only have solutions when the momentum of the leading mode in region II, $\kappa^{(2)}_0$, is imaginary, because otherwise the RHS would be imaginary while the LHS is real. This means that the plasmonic solutions are only allowed in the region $q > \sqrt{\vep_2} \omega/c$, what is not surprising, and is only a consequence of the bound nature of surface modes. On the other hand, the above equation can only have a solution when the tangent function present on its RHS is positive, what means that these solutions can only appear below a certain maximum frequency $\omega_{\ce{max}}$ given by
\begin{equation}
 \omega_{\ce{max}} =  \dfrac{\pi c}{2 h \sqrt{\vep_3}}.
 \label{eq:maxfreq}
\end{equation} 
Comparing this frequency with the effective plasma frequency given by equation \eqref{eq:plasmafreq}, we conclude that these are exactly the same, meaning that we recover the solution from the effective permittivity model with a completely different model, when we make the same approximation. Note that this expression has some limitations, namely the fact that it is only valid for frequency-independent permittivities, but nonetheless it is very useful to make a rough estimation of the order of magnitude of the plasmons frequency, although it tends to overestimate it; following the analogy to the perfect metal/dielectric interface, a better definition of a reference value for the spoof plasmons' fundamental mode frequency is $\omega_{{\ce{ref}}}=\omega_{\ce{max}}/\sqrt{1+\vep_1}$.

Apart from allowing the determination of $\omega_{\ce{max}}$, equation \eqref{eq:approxDR} has the additional advantage of being much easier to solve than the exact equation $\det(\mathbb{M}-\mathbb{I}) = 0$, which gets increasingly demanding with the number of modes we introduce in the fields description in region III.  Although this approximation looks somewhat naive, we will show later on that it produces really accurate results. %{\color{red} it is justified by the fact that, in such a narrow groove, the lifetime of the higher order terms should be much smaller than the leading order one}, and we will show later on that it produces really accurate results.

%Although this result is very satisfactory, we determined that the dispersion obtained through equation XXX strongly overestimates the energy of the spoof plasmons, which means that $\omega_{\ce{max}}$ can only be relied as a rough estimation of the order of magnitude of the plasmons frequency. A far better approximation is to consider that in region III (the grooves), only the $n=0$ mode exists, but everywhere else we allow the fields to be described by as many modes as necessary. Under these conditions, the dispersion relation of the spoof plasmons is given by the equation

\subsection{Optical Properties}
\label{sec:math:optical}

 If we add an impinging field in our description of the fields in region I [eq. \eqref{eq:BI}], the formalism described in the previous section allows for the calculation of the optical properties of the system (namely its reflectance). From a practical point of view, this result is particularly useful, because the reflectance of the system is easily measurable and allows the indirect measurement of other quantities like the absorbance spectrum and the identification of plasmonic resonances (note that there is no transmittance in this system, since a perfect metal is a perfect reflector).

 Assuming that the impinging electromagnetic wave has a frequency $\omega$ and makes some angle $\theta$ with the $y$-axis, the field in region I must be rewritten as
 \begin{equation}
 B_   {\ce{l}} \ccpar{x,y} = B_0 \sum_{n=-\infty}^{\infty} \ee^{\ii \beta_n x} \sqpar{ \ee^{-\ii k_y y} \delta_{n0} + r_n \ee^{\ii \kappa^{(1)}_n y} },
 \label{eq:BI:R}
 \end{equation}
 where $B_0$ is the intensity of the impinging magnetic field, and we have redefined the coefficients $B_n \equiv B_0 r_n$, so that the $r_n$ coefficients have the meaning of reflectance amplitudes. In the previous expression, $k_y$ is the momentum of the impinging wave in the $y$-direction, given by $k_y = k \cos\ccpar{\theta}$, $k=\sqrt{\vep_1} \omega/c$. On the other hand, Bloch's Theorem now imposes that $\beta_n = k_x + 2n\pi/d$, $k_x=k \sin\ccpar{\theta}$. From this point onwards, the procedure is completely analogous to the previous one. Following the already described steps (with the differences noted in the SI), we arrive at another matrix equation, this time with the form $(\mathbb{M} -\mathbb{I}) \cdot \mathbb{A} = \mathbb{F}$, where the matrices $\mathbb{M}$ and $\mathbb{I}$ are the same as before, $\mathbb{A}$ is a column with elements $[\mathbb{A}]_{\ell}=A_{\ell}$ and $\mathbb{F}$ is a column with elements $[\mathbb{F}]_{\ell} = \phi_{\ell}$, defined in equation \eqref{eq:phi}. 
 
 Unlike the previous case ---where we found the solution $(\mathbb{M} -\mathbb{I}) \cdot \mathbb{A} = 0$, with no source term---, the source introduced by the impinging wave allows the immediate determination of the coefficients $r_n$, upon the resolution of the previous equation [and using equation \eqref{eq:rfromA:R}]. From these coefficients, the reflectance of the system is simply defined as
 \begin{equation}
  \mathcal{R}\ccpar{\omega} = \sum_{n \in \mathrm{PM}} \mathrm{Re}\clpar{ \dfrac{\kappa^{(1)}_n}{\vep_1} } \mathrm{Re} \clpar{ \dfrac{\vep_1}{\kappa^{(1)}_0} } \sqabs{r_n}, 
 \end{equation} 
 where $\mathrm{Re}\clpar{x}$ stands for the real part of $x$, and both summations are performed strictly over the propagating modes (PM). For the energy scale of interest in this problem (up to a few THz), typically only the fundamental $n=0$ mode is propagating, and hence the reflectance takes the simpler form $\mathcal{R}\ccpar{\omega}=\sqabs{r_0}$. The absorbance, on the other hand, corresponds to the fraction of energy which is not reflected, being thus given by $\mathcal{A} = 1 - \mathcal{R}$.
 
 Besides enabling the calculation of the reflectance and absorbance spectra, the determination of the $r_n$ coefficients has the additional advantage of allowing the representation of the loss function\cite{goncalves2016introduction} of this problem, defined as $L(\omega,q) \equiv - \sum_n \ce{Im}\{r_n\}$ ($\ce{Im}\{x\}$ stands for the imaginary part of $x$), which allows for the indirect determination of the dispersion relation of the spoof plasmons ---even without the approximation considered in the previous section. Occasionally, we will be using a slightly different definition for the loss function, $\tilde{L}=\sign(L)\log(1+|L|)$, aimed at highlighting the dimmer dispersion curves in the presence of brighter ones. In the previous definition, $\sign(x)$ stands for the sign of $x$, and $\abs{x}$ stands for its absolute value.

 \section{Results and Discussion}
 \label{sec:results}

 %Although equation \eqref{eq:approxDR} is much simpler than the exact equation $\det(\mathbb{M}-\mathbb{I}) = 0$, it does not have an analytic solution, and thus needs to be evaluated numerically. This procedure is fairly simple under ideal conditions (eg., for spectrum-independent and real dielectric functions), but may get very demanding for more complicated cases. For that reason, in section \ref{sec:results:DR} we will only be considering the case where all regions I, II and III are filled with air (or vacuum), which will be useful to understand the general behaviour of the spoof plasmons, and the influence of the graphene sheet in its properties; in section \ref{sec:results:reflectance} we will be dealing with more complex and realistic systems, which will include several alcohols or water and hexagonal Boron-Nitride (hBN).
 
 For the results that will be presented henceforth, graphene's conductivity has been calculated through Mermin's non-local formula\cite{mermin1970lindhard,goncalves2016introduction}, synthetically presented in the SI. Moreover, although the results presented in the previous section were derived for isotropic media only, we will now occasionally consider hBN, which is anisotropic; the generalization of the previous results to this case is discussed in the SI as well. Finally, hBN's dielectric function was retrieved from Ref. \citenum{woessner2015highly}, Al$_2$O$_3$'s dielectric function was retrieved from Ref. \citenum{kischkat2012mid}, while the alcohols' (methanol, ethanol, 1-propanol and 2-propanol) and water's dielectric functions were retrieved from Ref. \citenum{li2014mesoscopic}.
 
 \subsection{Comparison to FDTD Simulations}
 \label{sec:results:FDTD}
 
 In the absence of graphene, equation \eqref{eq:approxDR} recovers the previously reported result for the dispersion relation of the spoof plasmons\cite{maradudin2014modern,maradudin2016rayleigh}. However, its applicability to describe real systems depends on the validity of the approximations employed so far, namely the fact that we have considered an ideal metal, and described the field inside the grooves with only one mode. To test the validity of these approximations, it is useful to compare the results we obtain analytically through equation \eqref{eq:approxDR} to fully-numerical results reported in the literature.
 
\begin{figure}
 \centering
 \includegraphics[width=0.75\textwidth]{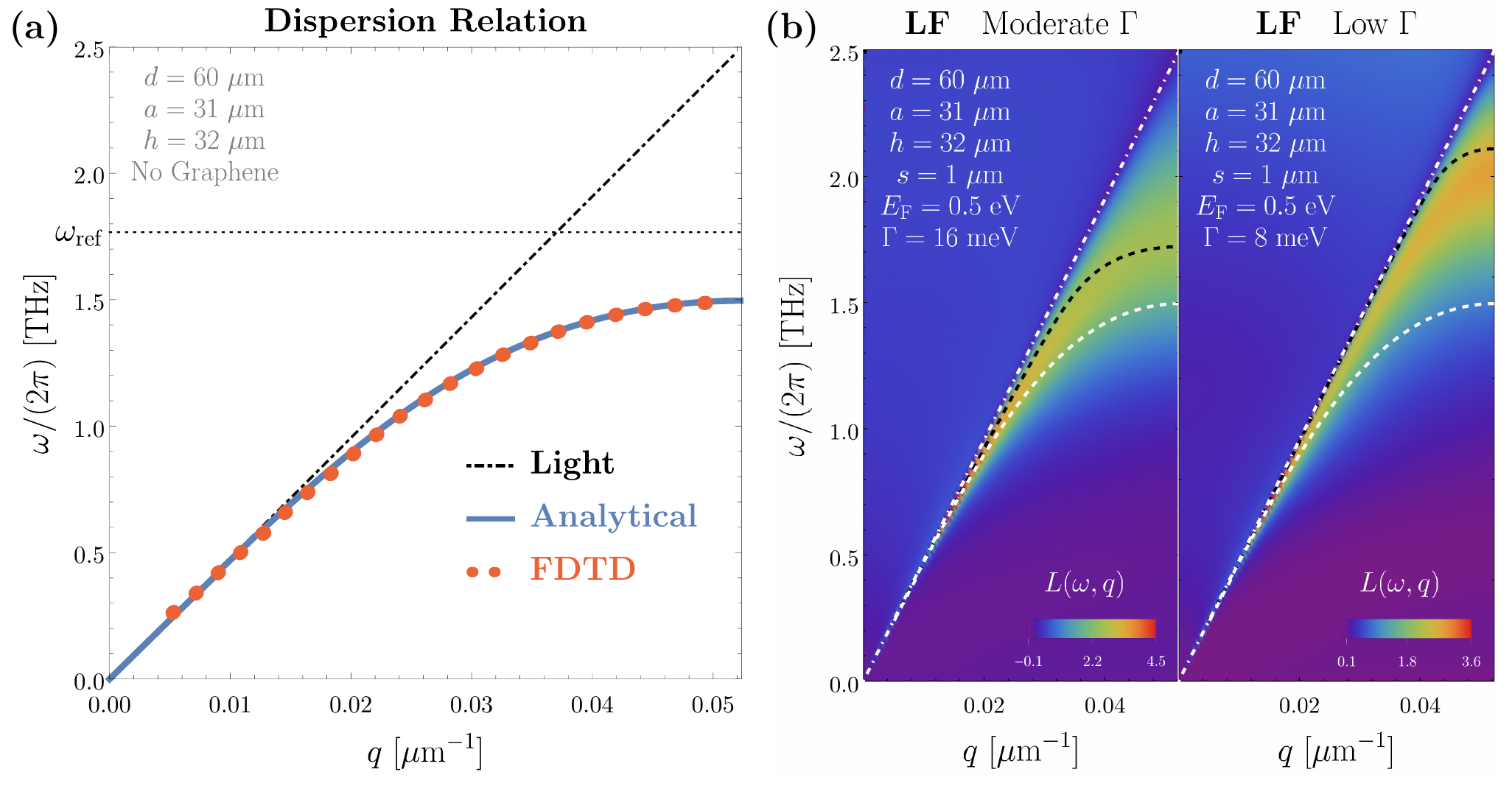}
 \caption{(a) Comparison between the dispersion relation of the spoof plasmons calculated through equation \eqref{eq:approxDR} (blue) and through an FDTD method (dotted red, retrieved from Ref. \citenum{ng2013spoof}), in the absence of graphene. (b) Loss function (LF) of the system in the presence of graphene, overlaid by the analytical dispersion relation of the same system with (dashed black) and without (dashed white) graphene, for two different values of the graphene damping energy $\Gamma$. All the parameters are specified in each plot. The dot-dashed line in each plot is the dispersion of the light.}
 \label{fig:FDTD}
 \end{figure}

 In particular, \citeauthor{ng2013spoof}\cite{ng2013spoof} studied a similar system to ours (albeit without graphene), and calculated the corresponding dispersion relation through an FDTD method using the commercial software \textit{Lumerical Solutions, Inc.}. Although the authors studied a system with trapezoidal grooves, instead of the rectangular grooves our model describes, we have considered an effective parameter for the width of those grooves (close to the average between the two trapeze bases), and calculated the dispersion relation for the same system. Both curves are plotted in Figure \ref{fig:FDTD}(a), and the agreement between them is excellent. This shows that the approximations described above do not jeopardize the utilization of this method to describe real systems.
 Moreover, Figure \ref{fig:FDTD}(a) also shows that $\omega_{{\ce{ref}}}$ provides indeed a good estimation of the order of magnitude of the spoof plasmons' fundamental mode frequency.
 
 In the presence of graphene, on the other hand, we can assess the validity of the single-mode approach inside the grooves by comparing the analytical solution in that approximation to the corresponding loss function spectrum (which considers an arbitrary number of modes). That study is represented in Figure \ref{fig:FDTD}(b) for two different damping regimes, where the black dashed curve is the analytical solution obtained though equation \eqref{eq:approxDR} for the indicated parameters, and the white dashed curve is the same as the blue one in Figure \ref{fig:FDTD}(a) (hence, without graphene). This Figure shows that the approximation carried out in equation \eqref{eq:approxDR} is very good for high-to-moderate values of the damping in the graphene; on the other hand, when the damping is low, the approximation is less accurate (it should ``detach'' from the light line at lower momenta), but nonetheless it provides a very good description of the maximum frequency of the SSP in the first Brillouin zone (measured at $q=\pi/d$). Furthermore, comparing the black and white dashed lines, Figure \ref{fig:FDTD}(b) suggests that the addition of the graphene tends to increase the energy of the spoof plasmons, what will be explored next.
 
 \subsection{Tuning of the Dispersion Relation}
 \label{sec:results:DR}
 
\begin{figure}
 \centering
 \includegraphics[width=0.75\textwidth]{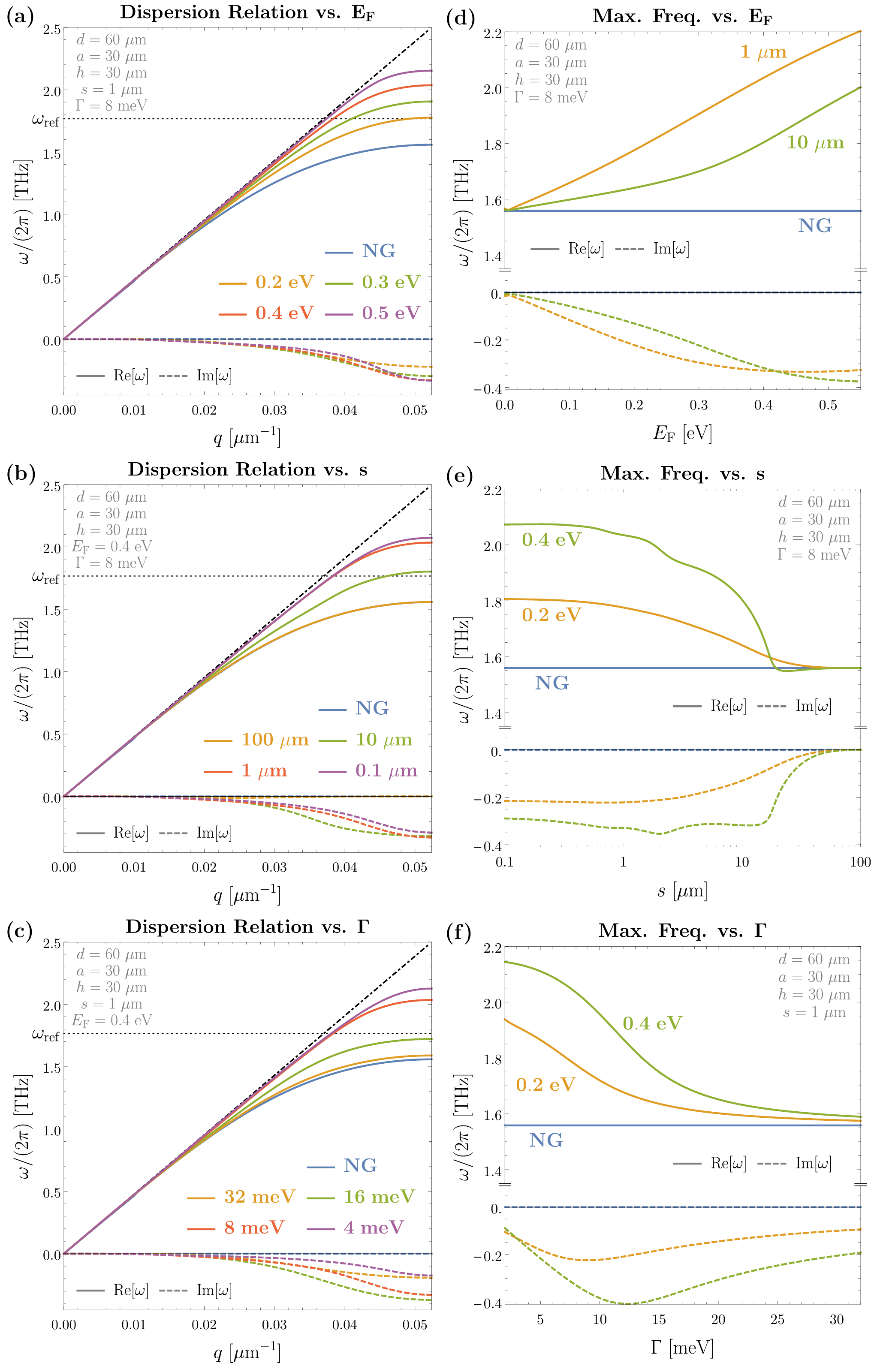}
 \caption{Top: dispersion relation of the spoof plasmons for several values of (a) the graphene's Fermi energy $E_{\ce{F}}$, (b) the spacer width $s$ and (c) the graphene's relaxation energy $\Gamma$. Bottom: maximum frequency of the plasmons (for $q=\pi/d$) in function of (a) $E_{\ce{F}}$, (b) $s$ and (c) $\Gamma$. All the parameters are specified in each plot. The dot-dashed line in the left-side plots is the dispersion of the light. `NG' stands for `no graphene'.}
 \label{fig:DR}
 \end{figure}

 The behaviour observed in Figure \ref{fig:FDTD}(b) suggests that the dispersion relation of the spoof plasmons can be tuned by the introduction of a graphene sheet in the system. This adds two additional parameters to the problem ---the graphene's Fermi energy $E_{\ce{F}}$ and the spacer width $s$--- that can be easily changed to control the energy of the spoof plasmons, while keeping the qualitative characteristics of their dispersion unchanged. This feature is shown in Figure \ref{fig:DR}, where are represented (a)--(c) the dispersion relation of the spoof plasmons when varying individually the graphene's Fermi energy $E_{\ce{F}}$, the spacer width $s$, and the graphene's relaxation energy $\Gamma$, respectively; and (d)--(f) the maximum frequency in the Brillouin zone (at $q=\pi/d$) in function of the same parameters.

 On the one hand, Figure \ref{fig:DR}(a) confirms that there is an actual scaling of the plasmons' energy due to the graphene, which can be higher than 40\%, for a doping up to $0.5 \unit{eV}$. This enhancement increases strictly (and almost linearly) with the increasing of $E_{\ce{F}}$, and its limit is settled by experimental limitations: in general, Fermi energies much greater than $0.5\unit{eV}$ are difficult to achieve\cite{efetov2010controlling,ye2011accessing}. On the other hand, this enhancement is greatly favoured by small spacer widths [see Figure\ref{fig:DR}(b)], what results from a stronger coupling between the metal and the graphene under those conditions. On the opposite regime (when the distance between the metal and the graphene increases considerably), these effectively decouple and we recover the `no graphene' behaviour. Nevertheless, Figure \ref{fig:DR}(e) shows that there is a slight saturation of the enhancement for spacers smaller than $\sim 0.5\unit{\mu m}$, what means that there is no significant gain in further reducing that dimension.
 
 Finally, Figures \ref{fig:DR}(c) and (f) intend to show that the relaxation energy of the graphene sheet plays a very important role in this analysis. Although this is not an actively changeable parameter, it should be relatively small in order to guarantee greater energy enhancements ---in fact, for $\Gamma$ values larger than $\sim 15 \unit{meV}$, the enhancement is very small even for high graphene doping and small spacer widths. Using hexagonal Boron Nitride as a spacer between graphene and the grating will reduce the value of $\Gamma$ significantly\cite{dean2010boron}.
 
\begin{figure}
 \centering
 \includegraphics[width=0.75\textwidth]{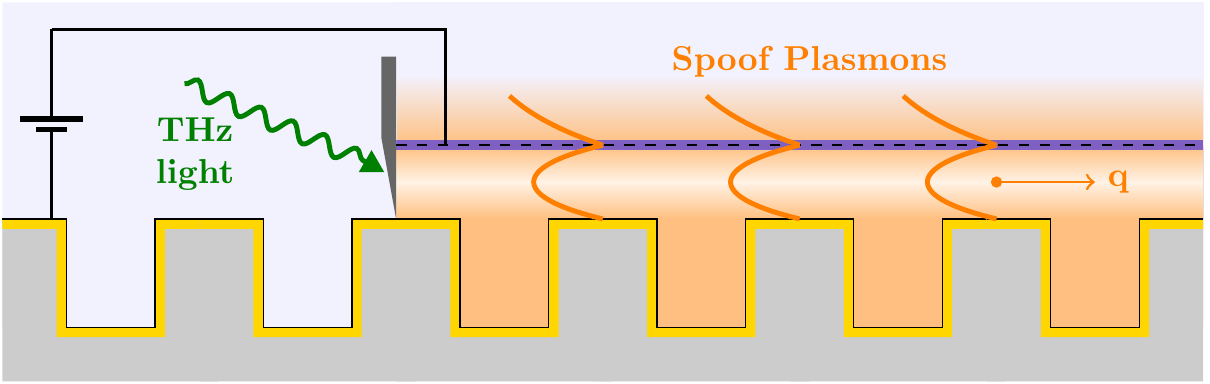}
 \caption{Example of a configuration that allows the excitation of spoof plasmons in the system. THz light is impinged in a thin metallic plate which only transmits evanescent modes. The momenta of the evanescent waves is higher than that of the incident light, being able to match the spoof plasmons momenta and thus excite them. The Fermi energy of the graphene sheet can be regulated by applying a variable gate potential.}
 \label{fig:Excitation}
 \end{figure}

 This behaviour may be of the utmost importance for actively controllable plasmonic waveguides in the THz spectral range. The spoof plasmons can be excited in a grooved surface using a system as depicted in Figure \ref{fig:Excitation}, which takes advantage of attenuated total reflection (ATR) method\cite{foley2015surface} to overcome the momentum mismatch between the impinging light and the bound surface modes (similarly to the well-known Kretschmann-Raether\cite{raether2006surface} or Otto\cite{otto1968excitation} configurations). Afterwards, one can tune the energy of these plasmons within a reasonable range by applying a gate voltage between the graphene sheet and the metal. Another advantage of using graphene is that its losses are very small: estimating the propagation length as $\zeta = 2 \pi v/\omega''$ ($v=\partial \omega'/\partial q$ is the group velocity of the spoof plasmons, with $\omega=\omega'-\ii \omega''$), our calculations predict that $\zeta$ is of the order of millimetres even when the graphene is highly doped ($E_{\ce{F}} \gtrsim 0.5\unit{eV}$); however, one must note that our model does not account for the damping in the metal itself. This is not a strong limitation, since in the THz the skin depth in the metal is very small ---thus validating our approach---, and hence losses in the metal will be small.
 
\begin{figure}
 \centering
 \includegraphics[width=0.75\textwidth]{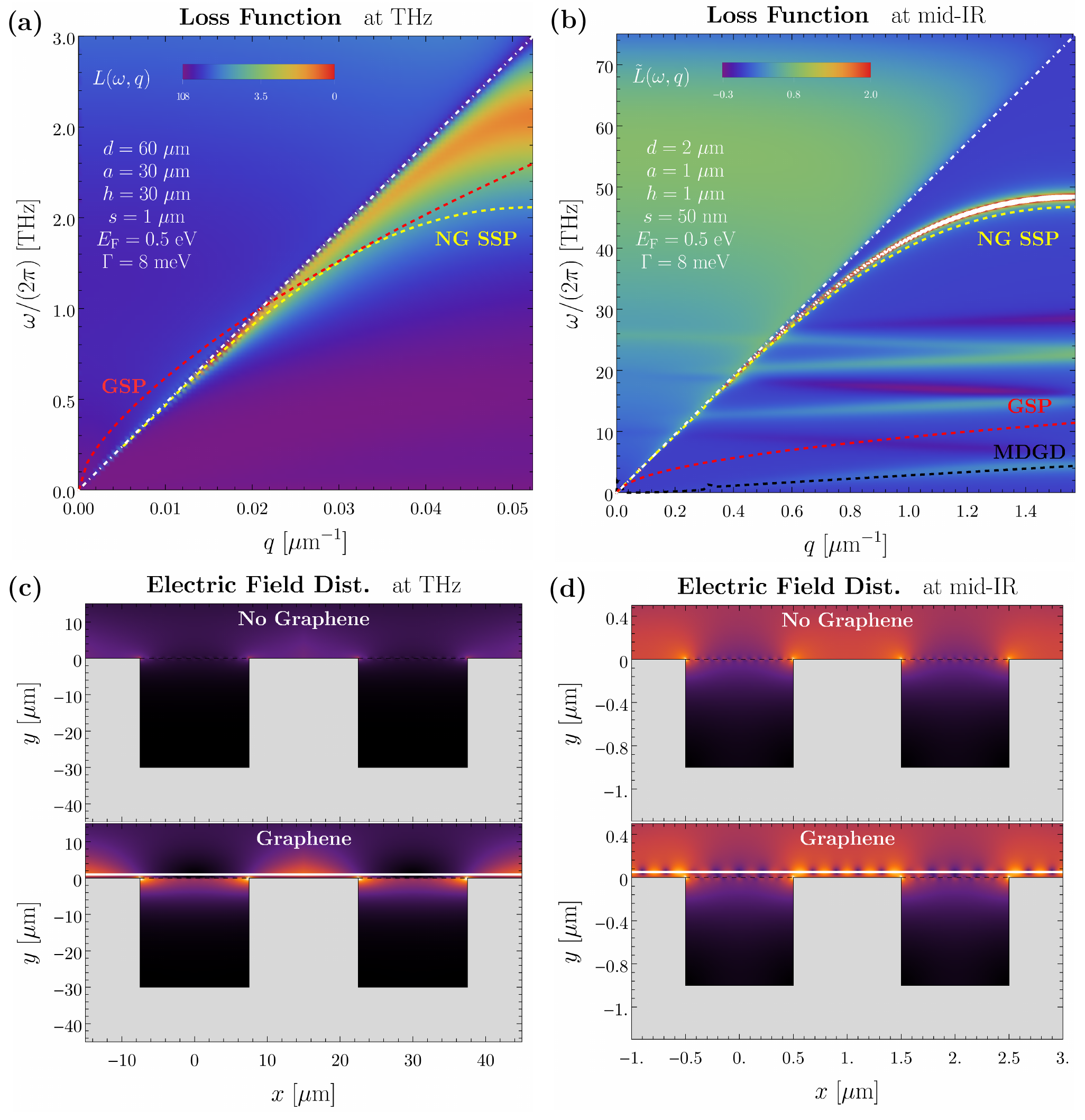}
 \caption{Top: Loss function spectrum of two different systems whose spoof plasmons lie on the (a) THz and (b) mid-IR spectral range, in the presence of doped graphene. Overlaid to the loss function are the dispersion of the spoof plasmons in the absence of graphene (`NG SSP', dashed yellow) and the dispersion of the graphene surface plasmons (`GSP', dashed red) in a air/graphene/air configuration. In (b) is also plotted the dispersion of the acoustic plasmons in a flat metal/air/graphene/air configuration (`MDGD', dashed black). The white dot-dashed curve is the dispersion of the light in the air. Bottom: electric field intensity in the vicinity of a groove for (c) the THz and (d) the mid-IR corresponding systems, with and without graphene. Both distributions were calculated for $q=0.9 \pi/d$ (for the corresponding $d$ in each system), what translates into the frequencies (c) $1.94\unit{THz}$ and $2.86\unit{THz}$ and (d) $46.3\unit{THz}$ and $47.3\unit{THz}$, respectively without and with graphene. All remaining parameters are the same as disclosed in (a) and (b). The color-scale is the same in both panels of each figure (c) or (d).}
 \label{fig:LossFunction:IR}
 \end{figure}

 Finally, it should be noted that the energy-enhanced spoof plasmons studied thus far are effectively hybrid modes that result from the coupling between the plasmons in the metal and in the graphene. Therefore, the strong enhancements observed in Figure \ref{fig:DR} are particular of the THz spectral range, and cannot, in general, be reproduced for much higher (or lower) energies. The reason for this behaviour is that, outside this spectral (THz) range, the characteristic energy scales of the plasmons on the graphene and the metal become very different, and they cannot efficiently couple. This feature is clearly visible in Figure \ref{fig:LossFunction:IR}, where are plotted side-by-side the loss functions of a system like the one studied in Figure \ref{fig:DR}, and one whose dimensions were reduced around 30-fold, what predicts a 30-fold increase in the frequency of the SSPs to around $\omega_{\ce{ref}}\sim 50\unit{THz}$, well deep in the mid-infrared (mid-IR) region. In both plots, the yellow dashed line corresponds to the dispersion of the spoof plasmons without graphene, and the red dashed line is the dispersions of the graphene surface plasmons (GSPs) in a air/graphene/air configuration (this configuration is used for simplicity of the analysis).
 
 It is clear that, in the THz range, these curves are very close, which translates into a strong coupling between the plasmons in the graphene and in the metal, and thus provokes a strong enhancement of the energy of the hybrid mode. In the mid-IR range, on the other hand, these curves have different energy scales and the plasmons do not couple efficiently, thus provoking a small enhancement of the hybrid mode energy. It is also interesting to note that, in the latter case, an additional low-energy mode arises in the spectrum, corresponding to a acoustic graphene plasmon in a flat metal/air/graphene/air configuration\cite{goncalves2016introduction} (black dashed line in the Figure). The Bragg reflections of this mode at the edges of the Brillouin zone are clearly visible. 
 
 Another evidence of this feature is visible in Figures \ref{fig:LossFunction:IR}(c) and (d), where the electric field intensity was plotted inside and in a vicinity of a groove for both systems above studied, and both in the presence and absence of graphene. While in (c) (THz range) the introduction of the graphene strongly changes the field distribution in the overall system (including inside the grooves) ---strongly enhancing the field in the groove wedges---, in Figure (d) (mid-IR) the changes are much less important, and concentrate only on the nearest vicinity of the graphene, where arise some field oscillations due to acoustic graphene plasmons.
 
 This analysis evidences that, in the smaller system, the plasmons in the metal and in the graphene are, in fact, decoupled, from where we conclude that graphene cannot be used to efficiently tune the energy of the SSPs in this spectral range.
 
 However, the fact that the graphene does not change the behaviour of the SSPs under these conditions may be useful for different applications ---for example, it has been shown that presence of the grating below the graphene provokes a strong enhancement of its optical absorption in the IR spectral range\cite{zhan2012band}. On a different perspective, graphene-coated metallic surfaces have been studied recently as an alternative to traditional metallic surfaces for optoelectronic applications, due to its much higher ability to resist oxidation and corrosion\cite{kravets2014graphene,ansell2015hybrid} (what is a recurring problem in plasmonics), while keeping (or even improving) the surface's characteristics.

\subsection{Application to filtering and sensing}  
  \label{sec:applications}
  
\begin{figure}
 \centering
 \includegraphics[width=0.75\textwidth]{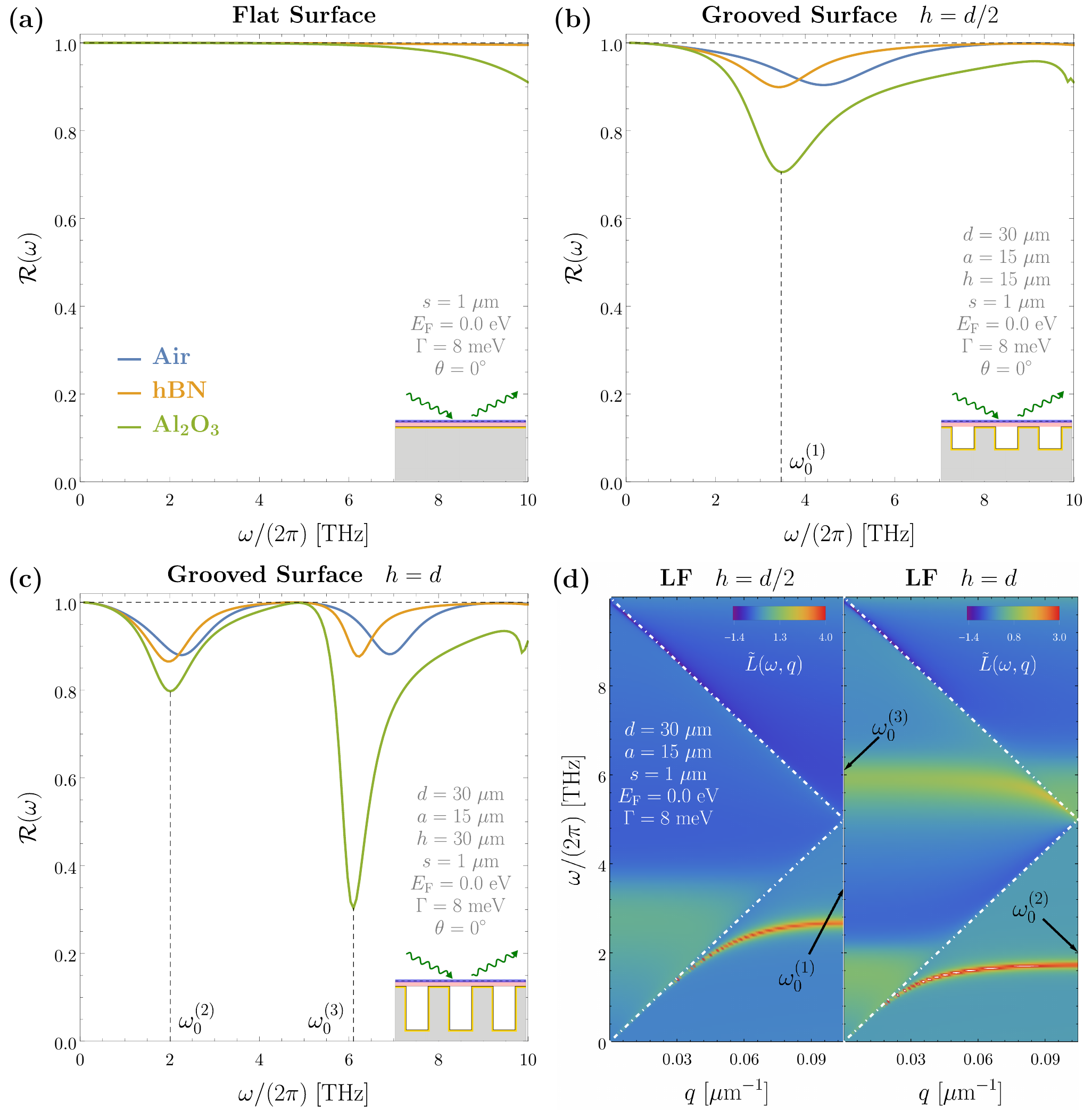}
 \caption{Reflectance spectrum of (a) a flat and (b),(c) a grooved metallic surface, with air above the graphene and inside the grooves, and several different materials in the spacer. Systems in plots (b) and (c) have different groove depths, what deeply influences the reflectance spectrum. In figure (d) is represented the loss function of the systems in (b) and (c) with Al$_2$O$_3$.}
 \label{fig:Reflectance:NoFluid}
 \end{figure}

 Coming back to the THz spectral range, spoof plasmons have a particularly interesting effect on the reflectance spectrum of the grating, which is visible in Figure \ref{fig:Reflectance:NoFluid}. In this figure, the reflectance spectra has been plotted for the case of a flat metal surface and a for two different geometries of a grooved metal. Comparing the three plots, one sees that the introduction of the grooves changes dramatically the reflectance spectrum of the system, which goes from an almost perfect reflector to exhibiting well-defined resonances corresponding to the excitation of spoof plasmons. These results were verified for several different materials in the spacer. Unsurprisingly, the position of these resonances is strongly controlled by the geometric dimensions of the system, since they are intrinsically connected to the dispersion relation of the spoof plasmons. This becomes clear in Figure \ref{fig:Reflectance:NoFluid}(d), where is represented the loss function of the air/Al$_2$O$_3$/air system. Comparing the position of the resonances in either case to the respective dispersion curves, one sees that, qualitatively, the resonance occurs where the dispersion intersects the Brillouin zone boundary.
 
 In the plots of the previous figure, a neutral graphene sheet has been placed in the system in order to add some damping which brightens the dispersion curves in Figure \ref{fig:Reflectance:NoFluid}(d); however, its conductivity is very low and therefore the behaviour of the system does not differ very much from the no-graphene case. It has the additional advantage of allowing the excitation of the spoof plasmons in the case where neither dielectric region is dispersive ---without the graphene, the reflectance spectrum of the air/air/air configuration would be identically equal to 1 even for the grooved system.
 
\begin{figure}
 \centering
 \includegraphics[width=0.75\textwidth]{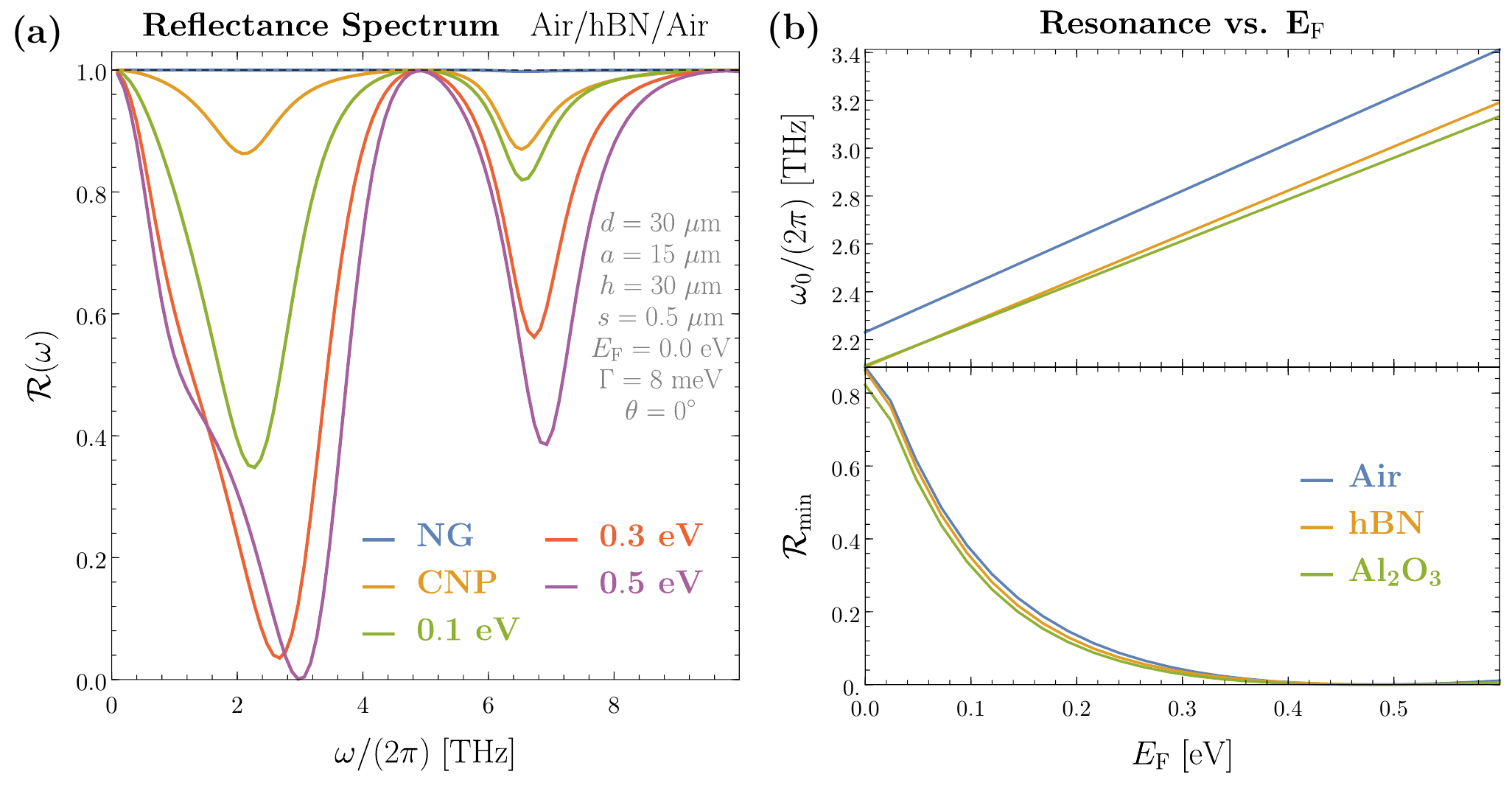}
 \caption{(a) Reflectance Spectrum of a graphene-coated metallic grating for several values of the Fermi energy of the graphene sheet. `NG' stands for `no graphene', whereas `CNP' stands for `charge neutral point', equivalent to $E_{\ce{F}}=0.0\unit{eV}$. (b) Position of the resonance (top) and minimum reflectance (bottom) of the reflectance spectrum, for several different spacer materials.}
 \label{fig:Reflectance:Resonance}
 \end{figure}

 However, a much more useful behaviour arises when the graphene is doped, as presented in Figure \ref{fig:Reflectance:Resonance}(a). Since the graphene doping strongly controls the dispersion curves of the system, it indirectly also controls the position of the resonances, which can be adjusted at will by varying the graphene's Fermi energy. Furthermore, the results show that the graphene sheet strongly increases of absorbance of the system even for low doping. This translates in very strong resonances, whose width is controlled by the graphene damping $\Gamma$, which can have their minimum as low as $\mathcal{R}=0$. This situation of full-absorbance can be very useful for the development of actively-tunnable filters in the THz spectral range. Figure \ref{fig:Reflectance:Resonance}(b) shows that this tuning may be superior than $1 \unit{THz}$, and for dopings above $0.3\unit{eV}$ the energy at the resonance is fully absorbed.

\begin{figure}
 \centering
 \includegraphics[width=0.75\textwidth]{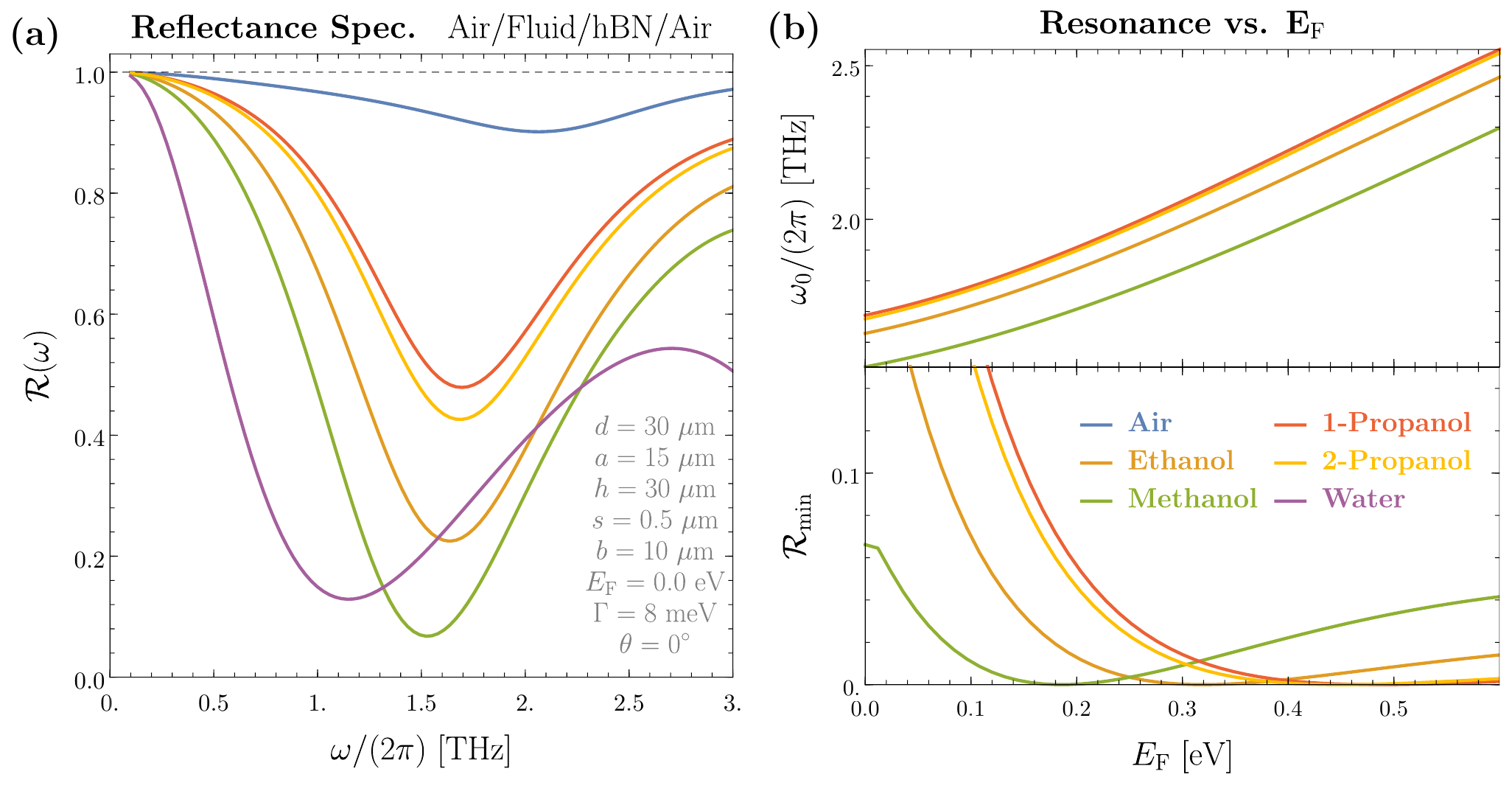}
 \caption{(a) Reflectance Spectrum of a graphene-coated metallic grating for several different fluids placed above the (neutral) graphene. (b) Position of the resonance (top) and minimum reflectance (bottom) of the reflectance spectrum, for the different alcohols being sensed.}
 \label{fig:Reflectance:Alcohols}
 \end{figure}

 On an alternative perspective, the difference between the spectra for different spacers in Figure \ref{fig:Reflectance:NoFluid} shows that the reflectance spectrum of the metal grating is highly sensitive to changes in the dielectric function of its composing materials, what suggests its utility for sensing applications. This approach has already been explored in the literature, with different authors proposing its usage to discern media with different refractive indexes\cite{ng2013spoof2,yao2014high} placed inside the grooves of a corrugated surface. In this work, however, we propose a different configuration in which the material to be sensed is placed above the graphene sheet in a layer with some thickness $b$. This approach is preferable when the aim is to sense very thin layers of fluids, in the order of $10\unit{\mu m}$. This thin layers are smaller than the wavelength of the THz radiation, so no Fabry-Perot oscillations occur, what poses a more challenging problem. In Figure \ref{fig:Reflectance:Alcohols}(a) is plotted the reflectance spectrum of an air/fluid/hBN/air configuration, where the fluid being sensed is either an alcohol (ethanol, methanol, 1,2-propanol), water, or none (air). 

This resonances visible in that spectrum, with neutral graphene, already allow to discern whether the material being sensed is water, an alcohol, or none, but the resonance position of the different alcohols is very similar, what may cause difficulties to discern them from each other. A possible solution to overcome this difficulty is to dope the graphene sheet and trace the resonance position and minimum reflectance with the graphene's Fermi energy, what is done in Figure \ref{fig:Reflectance:Alcohols}(b). On the one hand, the graphene doping increases the separation between the resonance position of the different alcohols up to 30\%; however, the greatest effect occurs in the minimum reflectance, since the total-absorption point for each alcohol arises for a different value of graphene doping, what provides an effective method to discern them. These results show how adding graphene to the system can also be used for sensing purposes.
 
 \section{Conclusions}
 
 In this work, we were able to show that the widely-studied tunnability of graphene can be successfully applied to a metallic grating in order to control its dispersion relation, especially in the THz spectral range. We also showed that this feature can have several applications, ranging from optoelectronic waveguides to filters and to THz sensing, with the additional bonus of providing
  graphene-protected metallic surfaces.
 
 The model we employed proved to be accurate when benchmarked against FDTD calculations, but has some limitations that should be properly emphasized. The most important one is the fact that the metal is considered ideal, which means that its validity is restricted to situations where the metal's skin depth is negligible, such as the case of the THz and mid-IR radiation in most good plasmonic metals. For the same reason, although this model accounts for non-local effects in the graphene, it cannot account for non-local effects in the metal surface, that should be important when the graphene sheet is very close (a few nanometers) from the metal surface; under those conditions, further corrections need to be employed to ensure experimentally accurate results\cite{luo2013surface,luo2014van}. However, in our case, the distance between graphene and the metal grating is large enough for 
 non-local effects to negligible compared to that length scale.

 \section*{Acknowledgments}
 
 E J Dias and N M R Peres acknowledge support from the European Commission through the project "Graphene-Driven Revolutions in ICT and Beyond" (Ref. No. 696656)
 and the Portuguese Foundation for Science and Technology (FCT) in the framework of the Strategic Financing UID/FIS/04650/2013. 

 \newpage
 
 \begin{suppinfo} 

 %\section{Appendices}
 
 \appendix
 
 \section{Calculation details of the dispersion relation of the spoof plasmons}
 \label{sec:ape:CalculationDetails}
 
 The coefficients $A_n$, $B_n$ and $C_n^{\pm}$ are determined using the boundary conditions across interfaces I/II and II/III. Starting by the former, these are the continuity of the tangential electric field, $E^x_   {\ce{l}}\ccpar{x,s}=E^x_   {\ce{ll}}\ccpar{x,s}$, and the discontinuity of the magnetic field proportional to the density of surface currents in the graphene sheet, $\vect{K}$, given by the expression $B_   {\ce{l}}\ccpar{x,s} - B_   {\ce{ll}}\ccpar{x,s} = \mu_0 \vect{K} \cdot \vers{x} $. According to Ohm's Law, the density of surface currents is proportional to the electric field in that surface, $\vect{K}\ccpar{\vect{r},q,\omega} = \sigma\ccpar{q,\omega} \vect{E}^{\parallel}\ccpar{\vect{r},q,\omega}$, where $\sigma\ccpar{q,\omega}$ is the surface's conductivity. This function is assumed to be uniform in every point of the sheet, but we allow it to be dependent on the momentum of the EM field, in order to account for non-local effects. Note that this discussion is valid for any 2D material in the interface between regions I and II (not only graphene, but also an electron gas or a doped 2D transition metal dichalcogenide, for example) as long as the adequate conductivity function for that material is considered.
 
 Since the electric fields in the neighbouring regions of this interface are composed by the superposition of several modes with different momenta, each mode that composes the total electric field is effectively influenced by a different conductivity, and the overall current density can be written as 
 \begin{equation}
  K^x(x) = \ccpar{\frac{\ii c^2}{\omega \vep_1}} \sum_{n=-\infty}^{\infty} \ii \kappa_n^{(1)} B_n \ee^{\ii \beta_n x} \sigma\ccpar{\beta_n, \omega} \ee^{\ii \kappa_n^{(1)} s}.  
 \end{equation} 
  Using the above expression and the mentioned boundary conditions, one finds that the  $C_n^{\pm}$ and the $B_n$ coefficients are related by the expression $C_n^{\pm} = \chi_n^{\pm} B_n$, with
 \begin{equation}
  \chi_n^{\pm} = \frac{1}{2} \sqpar{ 1 + \frac{ \sigma\ccpar{\beta_n, \omega} \kappa_n^{(1)}}{\omega \vep_0 \vep_1} \pm \frac{\vep_2 \kappa_n^{(1)}}{\vep_1 \kappa_n^{(2)}} } \ee^{\ii \kappa_n^{(1)} s} \ee^{ \mp \ii \kappa_n^{(2)} s}.
 \end{equation} 
 
 On the other hand, at the interface II/III, the tangential electric field is continuous, $E^x_   {\ce{ll}}\ccpar{x,0} = E^x_   {\ce{lll}}\ccpar{x,0}$, as well as the magnetic field, $B_   {\ce{ll}}\ccpar{x,0} = B_   {\ce{lll}}\ccpar{x,0}$, but only in the region $\abs{x}<a/2$; in the complementary region $a/2<\abs{x}<d/2$, the tangential electric field must vanish, $E^x_   {\ce{ll}}\ccpar{x,0} = 0$. Therefore, when multiplying the first condition by $\ee^{-\ii \beta_{\ell} x}$ and integrating it in $\abs{x}<a/2$, we can take advantage of the fact that the integrand must vanish at $a/2<\abs{x}<d/2$ to conclude that
 \begin{equation}
  \int_{-a/2}^{a/2} \dd{x} \ee^{-\ii \beta_{\ell} x} E^x_   {\ce{lll}}\ccpar{x,0} = \int_{-a/2}^{a/2} \dd{x} \ee^{-\ii \beta_{\ell} x} E^x_   {\ce{ll}}\ccpar{x,0} = \int_{-d/2}^{d/2} \dd{x} \ee^{-\ii \beta_{\ell} x} E^x_   {\ce{ll}}\ccpar{x,0}.
 \end{equation} 
  Further noting that the last term in the previous equation yields the integral $\int_{-d/2}^{d/2} \dd{x} \ee^{-\ii \beta_{\ell} x} \ee^{\ii \beta_{n} x} = d \delta_{\ell n}$, this boundary condition can be written as
  \begin{equation}
   B_{\ell} = \ii \frac{\vep_2}{\vep_3} \frac{a}{d} \ccpar{ \frac{1}{\chi_{\ell}^+ - \chi_{\ell}^-} } \sum_{n=0}^{\infty} \frac{\kappa_n^{(3)}}{\kappa_{\ell}^{(2)}} \sin\ccpar{ \kappa_n^{(3)} h } S_{\ell n} A_n,
  \label{eq:BfromA}
  \end{equation} 
  where the $S_{\ell n}$ is defined as the integral (with an analytical solution)
  \begin{equation}
   S_{\ell n} \equiv \frac{1}{a} \int_{-a/2}^{a/2} \dd{x} \ee^{-\ii \beta_{\ell} x} \cos\sqpar{\frac{n \pi}{a}\ccpar{x-\frac{a}{2}}}.
  \end{equation}
 
 The second boundary condition may, in turn, be multiplied by $\cos\sqpar{\frac{\ell \pi}{a}\ccpar{x-\frac{a}{2}}}$ and integrated it in $\abs{x}<a/2$, yielding the equation
 \begin{equation}
  A_{\ell} = \ccpar{\frac{2}{1+\delta_{\ell 0}}} \sum_{n=-\infty}^{\infty} \frac{S^{\ast}_{n\ell}}{\cos\ccpar{ \kappa_{\ell}^{(3)} h }} \ccpar{ \chi_n^+ + \chi_n^- } B_n.
  \label{eq:AfromB}
 \end{equation} 
  where we have used the integral $\int_{-d/2}^{d/2} \dd{x} \cos\sqpar{\frac{\ell \pi}{a}\ccpar{x-\frac{a}{2}}} \cos\sqpar{\frac{n \pi}{a} \ccpar{x-\frac{a}{2}}} = \frac{a}{2} \ccpar{1+\delta_{\ell 0}} \delta_{\ell n}$.
  
  Equations \eqref{eq:BfromA} and \eqref{eq:AfromB} relate reciprocally the coefficients $A_n$ and $B_n$; merging the two equations, we arrive at
  \begin{equation}
   A_{\ell} = \sum_{m=0}^{\infty} \clpar{ \ii \frac{\vep_2}{\vep_3} \frac{a}{d}  \ccpar{\frac{2}{1+\delta_{\ell 0}}}   \sum_{n=-\infty}^{\infty} \ccpar{ \frac{\chi_n^+ + \chi_n^-}{\chi_{n}^+ - \chi_{n}^-} } \frac{\kappa_m^{(3)}}{\kappa_{n}^{(2)}} \frac{\sin\ccpar{ \kappa_m^{(3)} h }}{\cos\ccpar{ \kappa_{\ell}^{(3)} h }} S_{n m} S^{\ast}_{n\ell} } A_m.
  \end{equation} 
  
  Defining the expression in brackets in the previous expression as $M_{\ell m}$, we may write it as $\sum_{m=0}^{\infty} \ccpar{ M_{\ell m} - \delta_{\ell m} } A_m = 0$, what may be rewritten in the matrix form
  \begin{equation}
   \mat{ M_{00} - 1 & M_{01} & \cdots \\ M_{10} & M_{11} - 1 & \cdots \\ \vdots & \vdots & \ddots } \cdot \mat{ A_0 \\ A_1 \\ \vdots } = \mat{0 \\ 0 \\ \vdots}.
  \end{equation} 
  
  It is now obvious that the previous equation can only admit solutions if the determinant of the square matrix in the LHS (designated hereby as $\mathbb{M}-\mathbb{I}$, where $\mathbb{I}$ is the unit matrix) vanishes. That equation, $\det\ccpar{ \mathbb{M} - \mathbb{I} } = 0$, is therefore the equation that sets the dispersion relation of the spoof plasmons allowed in this system.

 \section{Calculation details of the reflectance amplitudes}
 \label{sec:ape:ReflectanceAmplitudes}  
  
  The calculation procedure for this case is completely analogous to the previous one, using now the field in region I given by equation \eqref{eq:BI:R}. This slightly changes the solutions of the boundary conditions at the interface I/II, which now yield the equations $C_{\ell}^{\pm} = \chi_{\ell}^{\pm} r_{\ell} + \delta_{\ell 0} \Lambda^{\pm}$, with the $\chi_{\ell}^{\pm}$ being the same as before, and
  \begin{equation}
   \Lambda^{\pm} \equiv \frac{1}{2} \sqpar{ 1 - \frac{ \sigma\ccpar{\beta_n, \omega} k_y}{\omega \vep_0 \vep_1} \mp \frac{\vep_2 k_y}{\vep_1 \kappa_n^{(2)}} } \ee^{-\ii k_y s} \ee^{ \mp \ii \kappa_n^{(2)} s}.
  \end{equation} 
  
  Because of these new terms, the relations between the coefficients $A_m$ and $r_m$ are slightly changed to the equations
  \begin{equation}
   r_{\ell} = \ii \frac{\vep_2}{\vep_3} \frac{a}{d} \ccpar{ \frac{1}{\chi_{\ell}^+ - \chi_{\ell}^-} } \sum_{n=0}^{\infty} \frac{\kappa_n^{(3)}}{\kappa_{\ell}^{(2)}} \sin\ccpar{ \kappa_n^{(3)} h } S_{\ell n} A_n - \delta_{\ell 0} \ccpar{ \frac{\Lambda^+ - \Lambda^-}{\chi_0^+ - \chi_0^-} },
  \label{eq:rfromA:R}
  \end{equation} 
   \begin{equation}
  A_{\ell} = \ccpar{\frac{2}{1+\delta_{\ell 0}}} \sum_{n=-\infty}^{\infty} \frac{S^{\ast}_{n\ell}}{\cos ( \kappa_{\ell}^{(3)} h )} \ccpar{ \chi_n^+ + \chi_n^- } r_n + \ccpar{\frac{2}{1+\delta_{\ell 0}}} \frac{S^{\ast}_{0 \ell} \ccpar{ \Lambda^+ + \Lambda^- }}{\cos ( \kappa_{\ell}^{(3)} h )}.
  %A_{\ell} = \ccpar{\frac{2}{1+\delta_{\ell 0}}} \sum_{n=-\infty}^{\infty} \frac{S^{\ast}_{n\ell}}{\cos ( \kappa_{\ell}^{(3)} h )} \ccpar{ \chi_n^+ + \chi_n^- } r_n + \delta_{0 \ell} \sqpar{ \dfrac{ \Lambda^+ + \Lambda^- }{\cos ( \kappa_{\ell}^{(3)} h )}}.
  \label{eq:Afromr:R}
 \end{equation} 
  
  Merging once again the previous two equations, we arrive at an expression with the form $\sum_{m=0}^{\infty} \ccpar{ M_{\ell m} - \delta_{\ell m} } A_m = \phi_{\ell}$, or, in the matrix form,
  \begin{equation}
   \mat{ M_{00} - 1 & M_{01} & \cdots \\ M_{10} & M_{11} - 1 & \cdots \\ \vdots & \vdots & \ddots } \cdot \mat{ A_0 \\ A_1 \\ \vdots } = \mat{\phi_0 \\ \phi_1 \\ \vdots},
  \end{equation} 
  with
  \begin{equation}
   \phi_{\ell} \equiv - \ccpar{\frac{2}{1+\delta_{\ell 0}}} \frac{S^{\ast}_{0 \ell}}{\cos ( \kappa_{\ell}^{(3)} h ) } \clpar{ \Lambda^+ \sqpar{ 1 - \ccpar{ \frac{\chi_0^+ + \chi_0^-}{\chi_0^+ - \chi_0^-} } } + \Lambda^- \sqpar{ 1 + \ccpar{ \frac{\chi_0^+ + \chi_0^-}{\chi_0^+ - \chi_0^-} } } }.
   \label{eq:phi}
  \end{equation} 
  
  This is a readily solvable equation which allows the direct calculation of the $A_n$ coefficients, and the calculation of the $r_n$ coefficients using equation \eqref{eq:rfromA:R}.

 \section{Graphene's Conductivity}
 \label{sec:ape:Mermin}
 
 For the conductivity of the graphene sheet, we have used Mermin's formula, which includes non-local effects. Let $x \equiv q/k_{\ce{F}}$ and $y \equiv \hbar \omega/E_{\ce{F}}$ be dimensionless variables constructed from $q$ and $\omega$, respectively. $E_{\ce{F}}$ refers to the graphene's Fermi energy, $k_{\ce{F}} = E_{\ce{F}}/(\hbar v_{\ce{F}})$ is the Fermi momentum ($v_{\ce{F}} \approx c/300$ is the Fermi speed) and $\Gamma$ is the material's relaxation energy. The formula we have used was retrieved from \citeauthor{goncalves2016introduction}\cite{goncalves2016introduction},
  \begin{equation}
  \sigma\ccpar{q,\omega} = 4 \ii \sigma_0 \frac{\hbar \omega}{q^2} \chi_{\tau}\ccpar{\frac{q}{k_{\ce{F}}},\frac{\hbar \omega}{E_{\ce{F}}}},
 \end{equation} 
 with $\sigma_0 \equiv e^2 / (4\hbar)$ and
 \begin{equation}
  \chi_{\tau}\ccpar{x,y} = \frac{\ccpar{1+\ii \frac{\Gamma}{y E_{\ce{F}}}} \chi_{\ce{g}}\ccpar{x,y+\ii \frac{\Gamma}{E_{\ce{F}}}} }{1 + \ii \frac{\Gamma}{y E_{\ce{F}}} \chi_{\ce{g}}\ccpar{x,y+\ii \frac{\Gamma}{E_{\ce{F}}}} / \chi_{\ce{g}}\ccpar{x,0} }
 \end{equation} 
 
 \begin{equation}
  \chi_{\ce{g}}\ccpar{x,y} =
  \begin{cases}
   \chi_{\ce{B}}^{\ce{(1)}}\ccpar{x,y}, & \ce{Re}\sqpar{y}>x \wedge \ce{Re}\sqpar{y}<2-x, \\
   \chi_{\ce{B}}^{\ce{(2)}}\ccpar{x,y}, & \ce{Re}\sqpar{y}>x \wedge \ce{Re}\sqpar{y}>2-x, \\
   \chi_{\ce{B}}^{\ce{(3)}}\ccpar{x,y}, & \ce{Re}\sqpar{y}>2+x, \\
   \chi_{\ce{A}}^{\ce{(1)}}\ccpar{x,y}, & \ce{Re}\sqpar{y}<x \wedge \ce{Re}\sqpar{y}<2-x, \\
   \chi_{\ce{A}}^{\ce{(2)}}\ccpar{x,y}, & \ce{Re}\sqpar{y}<x \wedge \ce{Re}\sqpar{y}>2-x, \\
   \chi_{\ce{A}}^{\ce{(3)}}\ccpar{x,y}, & \ce{Re}\sqpar{y}<x-2.
  \end{cases}
 \end{equation} 
 
 \begin{align}
   \chi_{\ce{B}}^{\ce{(1)}}\ccpar{x,y} = -\frac{2}{\pi} \frac{E_{\ce{F}}}{(\hbar v_{\ce{F}})^2} + \frac{1}{4 \pi} \frac{E_{\ce{F}}}{(\hbar v_{\ce{F}})^2} \frac{x^2}{\sqrt{y^2 - x^2}} \sqpar{ F\ccpar{\frac{y+2}{x}} - F\ccpar{ \frac{2-y}{x} } },
   %\sqpar{ -\arccosh\ccpar{ \frac{y+2}{x} } + \right. \nonumber \\ \left. + \ccpar{\frac{y + 2}{x}} \sqrt{\left(\frac{y+2}{x} \right)^2-1} + \arccosh\ccpar{ \frac{2-y}{x} } - \ccpar{\frac{2-y}{x}} \sqrt{\left(\frac{2-y}{x} \right)^2-1} } 
 \end{align} 

  \begin{align}
   \chi_{\ce{B}}^{{\ce{(2)}}}\ccpar{x,y} = -\frac{2}{\pi} \frac{E_{\ce{F}}}{(\hbar v_{\ce{F}})^2} + \frac{1}{4 \pi} \frac{E_{\ce{F}}}{(\hbar v_{\ce{F}})^2} \frac{x^2}{\sqrt{y^2-x^2}} \sqpar{ F\ccpar{\frac{y+2}{x}} + \ii G\ccpar{\frac{2-y}{x}}},
   %\left[ -\arccosh\ccpar{\frac{y+2}{x}} + \right. \nonumber \\ \left. + \ccpar{\frac{y+2}{x}}\sqrt{\left(\frac{y+2}{x}\right)^2-1} - \ii \arccosh\ccpar{\frac{2-y}{x}} + \ii \ccpar{\frac{2-y}{x}}\sqrt{1-\left(\frac{2-y}{x}\right)^2} \right]
 \end{align} 

  \begin{align}
  \chi_{\ce{B}}^{\ce{(3)}}\ccpar{x,y} = -\frac{2}{\pi} \frac{E_{\ce{F}}}{(\hbar v_{\ce{F}})^2} + \frac{1}{4 \pi} \frac{E_{\ce{F}}}{(\hbar v_{\ce{F}})^2} \frac{x^2}{\sqrt{y^2-x^2}} \sqpar{ -\ii \pi + F\ccpar{\frac{y+2}{x}} -F\ccpar{\frac{y-2}{x}} },
 \end{align}
 
  \begin{align}
  \chi_{\ce{A}}^{\ce{(1)}}\ccpar{x,y} = -\frac{2}{\pi} \frac{E_{\ce{F}}}{(\hbar v_{\ce{F}})^2} - \frac{\ii}{4 \pi} \frac{E_{\ce{F}}}{(\hbar v_{\ce{F}})^2} \frac{x^2}{\sqrt{x^2 - y^2}} \sqpar{ F\ccpar{\frac{y+2}{x}} - F\ccpar{\frac{2-y}{x}} },
  %\sqpar{ \arccosh\ccpar{ \frac{y+2}{x} } - \right. \nonumber \\ \left. - \ccpar{\frac{y + 2}{x}} \sqrt{\left(\frac{y+2}{x} \right)^2-1} - \arccosh\ccpar{ \frac{2-y}{x} } + \ccpar{\frac{2-y}{x}} \sqrt{\left(\frac{2-y}{x} \right)^2-1} } 
 \end{align} 
 
   \begin{align}
  \chi_{\ce{A}}^{\ce{(2)}}\ccpar{x,y} = -\frac{2}{\pi} \frac{E_{\ce{F}}}{(\hbar v_{\ce{F}})^2} + \frac{\ii}{4 \pi} \frac{E_{\ce{F}}}{(\hbar v_{\ce{F}})^2} \frac{x^2}{\sqrt{x^2 - y^2}} \sqpar{\ii \pi - F\ccpar{\frac{y+2}{x}} + \ii G\ccpar{\frac{2-y}{x}} },
 \end{align} 
  \begin{equation}
   \chi_{\ce{A}}^{\ce{(3)}}\ccpar{x,y} = -\frac{2}{\pi} \frac{E_{\ce{F}}}{(\hbar v_{\ce{F}})^2} + \frac{1}{4 \pi} \frac{E_{\ce{F}}}{(\hbar v_{\ce{F}})^2} \frac{x^2}{\sqrt{x^2-y^2}} \sqpar{ -\pi + G\ccpar{\frac{y+2}{x}} - G\ccpar{\frac{y-2}{x}}},
 \end{equation} 
 where we defined $F(x) \equiv x \sqrt{x^2-1}-\arccosh\ccpar{x}$ and $G(x) \equiv x \sqrt{1-x^2}-\arccos\ccpar{x}$.
 
\section{Particular Case of Axial Materials}
\label{sec:ape:AxialMaterials}

Axial materials are a special case of isotropic materials characterized by the fact that they have different permittivities along ($\vep^y$) and perpendicularly ($\vep^x$) to their optical axis. An example of an axial material is the hexagonal Boron Nitride (hBN), whose optical axis we assume to be perpendicular to its surface and parallel to the $y$-axis.
 
As a consequence, its permittivity is described by a tensor rather than a scalar, what slightly changes the Maxwell equations which describe it. In particular, the electromagnetic wave equation takes the form $\vep^x (k_x^2+k_z^2) + \vep^y k_y^2 = \vep^x \vep^y \omega^2/c^2$, which means that equations \eqref{eq:BI}--\eqref{eq:BIII} still hold, with the only difference being that the correspondent momentum in region $\nu$ for the $n$th mode in the $y$ direction must be adapted as $\sqrt{\vep_{\nu} \omega^2/c^2 - \beta_n^2} \quad \to \quad \sqrt{\vep^x_{\nu} \omega^2/c^2 - (\vep^x_{\nu}/\vep^y_{\nu})\beta_n^2}$. Furthermore, the electric field is still related to the magnetic field through equation \eqref{eq:EfromB} with $\vep_{\nu} \to \vep_{\nu}^x$.
 
 With these two modifications, all the calculations presented in the previous section may be generalized almost effortlessly to the case in which any the regions of our system are filled with an axial dielectric medium.

 \end{suppinfo}
 
%\bibliography{References}
\providecommand{\latin}[1]{#1}
\makeatletter
\providecommand{\doi}
{\begingroup\let\do\@makeother\dospecials
	\catcode`\{=1 \catcode`\}=2 \doi@aux}
\providecommand{\doi@aux}[1]{\endgroup\texttt{#1}}
\makeatother
\providecommand*\mcitethebibliography{\thebibliography}
\csname @ifundefined\endcsname{endmcitethebibliography}
{\let\endmcitethebibliography\endthebibliography}{}

\end{document}